\documentstyle[twoside,fleqn,npb,epsfig,cite]{article}
\newcommand{\AmS}{{\protect\the\textfont2
  A\kern-.1667em\lower.5ex\hbox{M}\kern-.125emS}}

\input paperdef

\begin{document}

\thispagestyle{empty}
\setcounter{page}{0}
\def\thefootnote{\fnsymbol{footnote}}

\begin{flushright}
CERN--PH--TH/2004--115\\
hep-ph/0406245 \\
\date{\today}
\end{flushright}

\vspace{1cm}

\begin{center}

{\large\sc {\bf Precision SUSY Physics}}
\footnote{Invited talk given at {\em Loops \& Legs in Quantum Field
    Theory 2004}, Zinnowitz, Germany, April 2004}

\vspace{1cm}

{\sc S.~Heinemeyer$^{1\,}$%
\footnote{
email: Sven.Heinemeyer@cern.ch
}%
}

\vspace*{1cm}

$^1$ CERN TH division, Dept.\ of Physics, 1211 Geneva 23, Switzerland

\end{center}

\vspace*{1cm}

\begin{center}
{\bf Abstract}
\end{center}
We review the theoretical status and the future perspectives of the most
important electroweak precision observables in the MSSM. This
comprises the mass of the $W$~boson, $\MW$, the effective leptonic mixing
angle, $\sweff$, the mass of the lightest MSSM Higgs boson, $\mh$, and
the anomalous magnetic moment of the muon, $\amu$. 
The impact of the parametric
uncertainties from the experimental errors of the input parameters is
studied, and an estimate for the remaining uncertainties from unknown
higher-order corrections is given.
The need for future improvements in the theory predictions is
investigated. 

\def\thefootnote{\arabic{footnote}}
\setcounter{footnote}{0}

\newpage


\title{Precision SUSY Physics}

\author{S. Heinemeyer\address{CERN TH division, Dept.\ of Physics,
        1211 Geneva 23, Switzerland}}

\begin{abstract}
We review the theoretical status and the future perspectives of the most
important electroweak precision observables in the MSSM. This
comprises the mass of the $W$~boson, $\MW$, the effective leptonic mixing
angle, $\sweff$, the mass of the lightest MSSM Higgs boson, $\mh$, and
the anomalous magnetic moment of the muon, $\amu$. 
The impact of the parametric
uncertainties from the experimental errors of the input parameters is
studied, and an estimate for the remaining uncertainties from unknown
higher-order corrections is given.
The need for future improvements in the theory predictions is
investigated. 
\end{abstract}

\maketitle


\section{INTRODUCTION}

Theories based on Supersymmetry (SUSY) \cite{susy} are widely
considered as the theoretically most appealing extension of the
Standard Model (SM). 
SUSY predicts the existence of scalar partners to the SM fermions, 
and spin--1/2 partners to the gauge and Higgs bosons. So far, the
direct search for SUSY particles has not been successful, setting 
lower bounds of \order{100 \gev} on their masses~\cite{pdg}. 

An alternative way to probe SUSY is via the virtual effects of the 
additional particles to precision observables. 
This requires a very high precision 
of the experimental results as well as of the theoretical predictions.
The most relevant electroweak precision observables (EWPO) in this
context are the $W$~boson mass, $\MW$, the effective leptonic weak mixing
angle, $\sweff$, the mass of the lightest $\cp$-even MSSM Higgs boson,
$\mh$, and the anomalous magnetic moment of the muon, 
$\amu \equiv (g-2)_\mu$.

Concerning the EWPO's three different errors have to be
distinguished: 

1.~\underline{the experimental error:} 
the future anticipated accuracy sets the scale
that has to be matched with the other two types of errors.

2.~\underline{the intrinsic error:} 
this error is due to unknown higher-order corrections. 
We will review the current status in the theoretical prediction of
the EWPO's. 
Emphasis is being put on the missing calculations to
match the future experimental error.

3.~\underline{the parametric error:} 
experimental errors in the input parameters
yield this uncertainty in the prediction of the EWPO's. In principle
this applies to the SM as well as to the SUSY parameters. However, the
future uncertainty in the SUSY parameters are highly model
dependent. Therefore we will not investigate their impact here.
The status and the future expectation of the SM parametric errors is
being analyzed.


Provided a high accuracy in both, the experimental determination and
the theoretical predictions for the EWPO's, 
electroweak precision tests (i.e.\ the comparison of accurate
measurements with predictions of the theory), allow 
to set indirect constraints on unknown parameters of the MSSM.
SUSY higher-order corrections to $\MW$, $\sweff$ and $\mh$ depend
most strongly on the third generation scalar quarks. $\amu$ depends at the
one-loop level 
on the second generation sleptons and on the masses of the neutralinos
and charginos. Thus precise measurements of these observables allow
to obtain indirect information on various parts of the SUSY spectrum.


\section{THE $W$ BOSON MASS AND THE EFFECTIVE MIXING ANGLE}

The two most common electroweak precision observables (EWPO) that are
used to check the validity of the SM or the MSSM are the $\MW$ and
$\sweff$. 
$\MW$ can be obtained iteratively from 
\BE
\MW^2 \KL 1 - \frac{\MW^2}{\MZ^2} \KR = \frac{\pi\,\al}{\wz\,\gf}\,
                                        \ed{1 - \De r}~,
\end{equation}
where $\De r$ contains the higher-order corrections.
The effective weak leptonic mixing angle is defined as
\BE
\sweff = 1/(4\,|Q_f|) \, (1 - \re g_V^f/\re g_A^f)~,
\end{equation}
where $g_{V,A}^f$ are the couplings of a fermion $f$ to the $Z$~boson
on the $Z$~resonance, $Q_f$ is the corresponding electric coupling,
and higher-order contributions enter through corrections to
$g_{V,A}^f$. 
The status and the future expectations of the three errors is as follows:

1.~The current and anticipated future experimental uncertainties are
summarized in \refta{tab:ewpofut}. 
See \cite{blueband} for a detailed discussion and further references.

2.~The SM part%
\footnote{
For sake of brevity we omit the references to SM calculations and
refer to \cite{susyewpo} and references therein.
}%
~of the MSSM evaluation of $\MW$ and $\sweff$ is
quite advanced, leading to~\cite{ACFW03}
\BE
\de\MW^{\SM} \approx 4 \mev, \; 
\de\sweff^{\SM} \approx 6 \cdot 10^{-5}
\end{equation}
The full one-loop contributions to $\MW$ and $\sweff$
arising in the MSSM can be found in \cite{dr1lB}. The leading
two-loop corrections, entering via the $\rho$~parameter, 
have been obtained at \order{\al\als}~\cite{dr2lA} and
\order{\alt^2,\alt\alb,\alb^2}~\cite{dr2lal2}. The leading gluonic
corrections to $\De r$ of \order{\al\als} (i.e.\ the only two-loop
calculation beyond the $\De\rho$ approximation) has been obtained in
\cite{dr2lB}. Using the methods described in \cite{blueband} we
arrive at the estimate~\cite{susyewpo}
\BE
\de\MW^{\rm MSSM} \approx 10 \mev, \; 
\de\sweff^{\rm MSSM} \approx 12 \cdot 10^{-5}
\end{equation}

3.~The most important parametric errors for $\MW$ and $\sweff$ come from
the $\mt$ and the hadronic contribution to the fine structure
constant, $\De\al_{\rm had}$. Currently we have
\BEA
\lefteqn{\de\mt = 4.3 \gev \Rightarrow} \\ 
&& \mbox{}\hspace{-5mm} \de\MW^{\mt} \approx 26 \mev, \; 
   \de\sweff^{\mt} \approx 14 \cdot 10^{-5} \non \\
\lefteqn{\de(\De\al_{\rm had}) = 36 \cdot 10^{-5} \Rightarrow} \\
&& \mbox{}\hspace{-5mm} \de\MW^{\De\al} \approx 6.5 \mev, \; 
   \de\sweff^{\De\al} \approx 13 \cdot 10^{-5} \non 
\EEA
For the future one can hope for
\BEA
\lefteqn{\de\mt = 0.1 \gev \mbox{\cite{tesla}} \Rightarrow} \\ 
&& \mbox{}\hspace{-5mm} \de\MW^{\mt} \approx 1 \mev, \; 
   \de\sweff^{\mt} \approx 0.3 \cdot 10^{-5} \non \\
\lefteqn{\de(\De\al_{\rm had}) = 5 \cdot 10^{-5} 
                               \mbox{\cite{Dealhadfut}} \Rightarrow} \\
&& \mbox{}\hspace{-5mm} \de\MW^{\De\al} \approx 1 \mev, \; 
   \de\sweff^{\De\al} \approx 1.8 \cdot 10^{-5} \non
\EEA
%
\begin{table}[tb!]
\caption{Uncertainties of $\sweff$ and $\MW$.
See \cite{blueband} for a detailed discussion and further references.}
\label{tab:ewpofut}
\renewcommand{\arraystretch}{1.1}
\BC
\begin{tabular}{|c||c|c|}
\cline{2-3} \multicolumn{1}{c||}{}
& $\de\MW$ [MeV] & $\de\sweff [10^{-5}]$ \\
\hline\hline
now & 34 & 17 \\ \hline
LHC & 15 & 14--20 \\ \hline
LC  & 10 &  -- \\ \hline
GigaZ & 7 & 1.3 \\ \hline
\end{tabular}
\EC
\renewcommand{\arraystretch}{1}
\vspace{-8mm}
\end{table}
%
By comparing the LC/GigaZ error with the future
parametric error it can be seen that $\MW$ will be well under
control. However, even with the optimistic assumption for 
$\De\al_{\rm had}$ the experimental GigaZ precision can hardly be
matched. Concerning the intrinsic error, especially for $\sweff$, a
large effort, probably a full two-loop calculation, will be necessary
to arrive at the required GigaZ precision.


\section{THE LIGHT MSSM HIGGS MASS}

The mass of the lightest $\cp$-even MSSM Higgs boson can be predicted from 
the other model parameters. At the tree-level, the two $\cp$-even Higgs 
boson masses are obtained as a function of $\MZ$, the $\cp$-odd Higgs
boson mass $\MA$, and the ratio of the two vacuum expectation
values $\tb$. 
In the Feynman-diagrammatic (FD) approach the higher-order corrected 
Higgs boson masses are derived by finding the
poles of the $h,H$-propagator 
matrix. This is equivalent to solving 
\BEA
\label{eq:proppole}
&0&= \left[p^2 - m_{h, \rm tree}^2 + \hSi_{hh}(p^2) \right] \times \\
&& \left[p^2 - m_{H, \rm tree}^2 + \hSi_{HH}(p^2) \right] 
- \left[\hSi_{hH}(p^2)\right]^2 ~. \non
\EEA
where the $\hSi(p^2)$ denote the renormalized Higgs-boson self-energies,
$p$ is the external momentum.

The status of the available results for the self-energy contributions to
\refeq{eq:proppole} in the real MSSM can be summarized as follows. For the
one-loop part, the complete result within the MSSM is 
known~\cite{ERZ,mhiggsf1lB,mhiggsf1lC}. 
Concerning the two-loop
effects, their computation is quite advanced, see \cite{mhiggsAEC} and
references therein. They include the strong corrections
at \order{\al_t\als}, and Yukawa corrections, \order{\al_t^2},
to the dominant one-loop \order{\al_t} term, as well as the strong
corrections from the bottom/sbottom sector at \order{\al_b\als}. 
For the $b/\Sbot$~sector
corrections also an all-order resummation of the $\Tb$-enhanced terms,
\order{\al_b(\als\tb)^n}, is known~\cite{deltamb}.
Most recently the \order{\al_t \al_b} and \order{\al_b^2} corrections
have been derived~\cite{mhiggsEP5}.%
\footnote{
Leading corrections in the MSSM with non-minimal flavor violation 
have recently been obtained in~\cite{mhiggsNMFV}.
}%

An upper bound of $\mh \lsim 140 \gev$~\cite{mhiggslong,mhiggsAEC} can
be established~\cite{feynhiggs} taking into 
account all existing higher-order corrections 
($\mhmax$~scenario, $\mt = 178.0 \gev$, $\msusy = 1 \tev$), 
neglecting intrinsic uncertainties.
The status and the future expectations of the three errors is as follows:


1.~The experimental error will be $\De\mh^{\rm exp} \approx 200 \mev$
at the LHC, provided the channel $gg \to h \to \ga\ga$ is sufficiently
strong~\cite{lhc}. At the LC a mass determination down to 
$\De\mh^{\rm exp} \approx 50 \mev$ will be possible~\cite{tesla}.

2.~The current intrinsic error consists of four different pieces:\\
$-$ missing momentum-independent two-loop corrections: By varying the
renormalization scale at the one-loop level, these two-loop
uncertainties can be estimated to be 
$\pm 1.5 \gev$~\cite{feynhiggs1.2}.%
\footnote{We do not consider here the
``full'' two-loop effective potential calculation presented in
\cite{fullEP2l}, since they have been obtained in a special,
simplified renormalization that make them unusable for the FD
approach.
}%

\noindent
$-$ missing momentum-dependent two-loop corrections: since at the
one-loop level the momentum corrections are below the level of 
$2 \gev$, it can be estimated that they stay below 
$\pm 0.5 \gev$~\cite{mhiggsAEC}.\\
$-$ missing 3/4-loop corrections from the $t/\Stop$~sector: by
applying three different methods (changing the renormalization scheme 
at the two-loop level; direct evaluation of the leading terms in a
simplified approximation; numerical iterative solution of the
renormalization group equations) these corrections have been estimated
to be at about $\pm 1.5 \gev$ (see \cite{mhiggsAEC} and
references therein). \\
$-$ missing 3/4-loop corrections from the $b/\Sbot$~sector:
the corrections from the $b/\Sbot$~sector can be large if both, $\mu$
and $\tb$ are sufficiently large. For $\mu > 0$ it can been
shown~\cite{mhiggsFD2} that 
the two-loop corrections give already an extremely precise result,
provided that the resummation of $(\als\tb)^n$ terms~\cite{deltamb} is
taken into 
account. On the other hand, for $\mu < 0$ the 3-loop
corrections can be up to $\pm 3 \gev$~\cite{mhiggsFD2}. Since the
results for $\amu$ favor a positive $\mu$ (see below) we do not
consider this possibility here.\\
The current intrinsic error can thus be estimated to be 
$\pm 3 \gev$~\cite{mhiggsAEC}. 

If the full two-loop calculation (in an FD suitable renormalization)
as well as the leading 3-loop (and possibly the very leading 4-loop)
corrections are available, the intrinsic error could be reduced to
about $\pm 0.5 \gev$. This seems to be possible within the next
5--10 years.

3.~The currently induced error by $\MW$ and $\mb$ are already almost
negligible, and will be irrelevant with the future precision of these
input parameters~\cite{deltamt}. On the other hand, $\mt$ and $\als$ play a
non-negligible role. Currently we have~\cite{deltamt}
\BEA
\de\mt \approx 4.3 \gev 
&\Rightarrow&
\de\mh^{\mt} \approx 4 \gev \\
\de\als \approx 0.002
&\Rightarrow&
\de\mh^{\als} \approx 0.3 \gev
\EEA
From the LC one can hope to achieve in the future
\BEA
\de\mt \approx 0.1 \gev 
&\Rightarrow&
\de\mh^{\mt} \approx 0.1 \gev \\
\de\als \lsim 0.001
&\Rightarrow&
\de\mh^{\als} \approx 0.1 \gev
\EEA


By comparing the LC (or even the LHC) precision for $\mh$ with the
intrinsic and parametric error, it becomes clear that a huge effort
from both the theoretical and from the experimental side will
be necessary in order to fully exploit the precise $\mh$
measurement. Without a reduction of the intrinsic error by about a
factor of~10, even the LHC precision will be worthless. The parametric
uncertainty emphasizes the complementarity of the LHC and the
LC. Already for the LHC precision of $\mh$ the LC precision of $\mt$
will be needed in order to match the level of $\de\mh \approx 200 \mev$.


\section{THE ANOMALOUS MAGNETIC MOMENT OF THE MUON}

The final result of the Brookhaven ``Muon $g-2$ Experiment'' (E821) for
the anomalous magnetic moment of the muon, $\amu \equiv (g-2)_\mu/2$, 
reads \cite{g-2exp}
\BE
\amuexp = (11\, 659\, 208 \pm 6) \times 10^{-10}~.
\label{eq:amuexp}
\EE
It is unclear whether this result will be improved within the next
$\sim$10 years. 
The SM prediction depends on the evaluation of the
hadronic vacuum polarization \cite{DEHZ,g-2HMNT,Jegerlehner,Yndurain},
and light-by-light contributions~\cite{LBL} (for a recent
reevaluation describing a possible shift of the central
value by $5.6 \times 10^{-10}$, see~\cite{LBLnew}).
Depending on the hadronic evaluation the
difference between experiment and the SM prediction lies between%
\footnote{
These evaluations are all $e^+e^-$ data driven. 
Recent analyses concerning $\tau$ data indicate that uncertainties due to
isospin breaking effects may have been underestimated
earlier~\cite{Jegerlehner}.
}%
\BEA
&& \mbox{}\hspace{-5mm} \amuexp-\amutheo = \non \\
\label{deviation1}
&& \mbox{}\hspace{-5mm} \mbox{\cite{g-2HMNT}+\cite{LBL}}:~ 
   (31.7\pm9.5)\times10^{-10} \,:\,3.3\,\si~, \\
&& \mbox{}\hspace{-5mm} \mbox{\cite{DEHZ}+\cite{LBLnew}}:~
   (20.2\pm9.0)\times10^{-10} \,:\,2.1\,\si~. 
\label{deviation2}
\EEA
There is hope that the comparison of the SM prediction with
the experimental result can become much more precise, even without new
direct experimental data on $\amu$. This will require on the one hand
a better understanding of the light-by-light contributions, and on the
other hand a better control (driven by new experimental data) on the
hadronic corrections to $\amu$. 

This discrepancy between experiment and SM prediction can be easily
explained by SUSY.  
The supersymmetric one-loop contribution \cite{g-2MSSMf1l}
is approximately given by 
\BE
\amu^{\SU,{\rm 1L}} \approx 13 \times 10^{-10} 
             \KL \frac{100 \gev}{\msusy} \KR^2 \tb~,
\label{susy1loop}
\EE
if all SUSY particles (here smuon,
sneutrino, chargino and neutralino) have a common mass
$\msusy$, and $\mu > 0$. 
Obviously, supersymmetric effects can easily account for a
$(20\ldots30)\times10^{-10}$ deviation, if 
$\msusy$ lies roughly between 100 GeV (for small $\tb$) and
600 GeV (for large $\tb$). 
Eq.~(\ref{susy1loop}) also shows that for certain parameter choices
the supersymmetric contributions could lie outside the $3\si$ band of
the allowed range according to (\ref{deviation1}),
(\ref{deviation2}). This means that the $(g-2)_\mu$ measurement places
strong bounds on the SUSY
parameter space. This is important for constraining different variants
of SUSY models and complements the direct searches. 
Even after
the discovery of supersymmetric particles, indirect bounds derived from 
$(g-2)_\mu$ will provide important complementary information to that
obtained from direct measurements. 

In order to fully exploit the precision of the $(g-2)_\mu$ experiment 
within SUSY, a reduction of the intrinsic error down to the level of
about $\pm1\times10^{-10}$ is 
desirable. This level has been reached for the perturbative part of
the SM evaluation, see \cite{g-2review} and references therein.

For the SUSY contributions, a similar level of accuracy has not been
reached yet, since the the status of the corresponding
two-loop corrections is much less advanced. 
Only four parts of the two-loop contribution have been
evaluated up to now. The
first part are the leading $\log \KL m_\mu/\msusy\KR$-terms of
SUSY one-loop diagrams with a photon in the second loop. They
amount to about $-8\%$ of the supersymmetric one-loop contribution
(for a SUSY mass scale of $\msusy = 500
\gev$)~\cite{g-2MSSMlog2l}. 

The second known part are the diagrams
with a closed loop of SM fermions or scalar fermions calculated in
\cite{g-2FSf}.
It has been shown in \cite{g-2FSf} that, if all experimental
constraints are taken into account, the numerical effect of these
contributions amount up
to about $5\times10^{-10}$, except in rather restricted parameter
regions with non-universal sfermion mass parameters involving very
disparate mass scales.

The third part consists of diagrams with a closed chargino/neutralino
loop, evaluated in \cite{g-2CNH}. These corrections are suppressed
by a factor of $\sim 50$ compared to the one-loop result if all SUSY
masses have roughly the same value. However, if the one-loop result is
suppressed by heavy slepton masses, the two-loop corrections can be of
the same order. In general they can amount up to $\sim 5 \times 10^{-10}$.

The fourth part are the diagrams that arise from the electroweak
two-Higgs-doublet model part of the MSSM. They also have been evaluated in
\cite{g-2CNH}. These contributions are in general small as compared to
the one-loop result and hardly exceed $2 \times 10^{-10}$. 

Despite the recent progress in the evaluation of two-loop corrections,
the remaining uncertainties are still larger than the
required $1 \times 10^{-10}$. The missing SM/SUSY corrections to the one-loop
MSSM result can be enhanced by large top and bottom Yukawa couplings; 
the two-loop QED corrections~\cite{g-2MSSMlog2l} could be modified if
the SUSY particles do not have one common mass scale. 
If the full two-loop result
will be available the intrinsic uncertainties for $\amu^{\SU}$ will be
reduced to the required level, provided that the mass scales in the
MSSM are not extremely disparate (which will be tested
experimentally).


\section*{Acknowledgements}
I thank the organizers of L \& L 2004 for the kind invitation and the
stimulating atmosphere.



\end{document}